%% file: main_taes.tex
\documentclass{IEEEtaes}

\usepackage{color,array,amsthm}
\usepackage{graphicx}

\jvol{XX}
\jnum{XX}
\jmonth{XXXXX}
\paper{1234567}
\pubyear{2022}
\doiinfo{TAES.2022.Doi Number}

\usepackage{algpseudocode}
\usepackage{subfig}
\usepackage{url}
\usepackage{cite}
\usepackage{siunitx}
\usepackage{booktabs}
\usepackage{multirow}

\usepackage[shortcuts,acronym]{glossaries}

\newacronym{gospa}{GOSPA}{Generalized Optimal Subpattern Assignment}
\newacronym{t2ta}{T2TA}{Track-to-Track Association}
\newacronym{t2tf}{T2TF}{Track-to-Track Fusion}
\newacronym{mot}{MOT}{Multi-Object Tracking}
\newacronym{rsu}{RSU}{Road-Side Unit}
\newacronym{so}{SO}{Stochastic Optimization}
\newacronym{so-c}{SO-C}{SO-Const.}
\newacronym{so-cs}{SO-CS}{SO-Const.-per-Sensor}
\newacronym{so-ds}{SO-DS}{SO-Distance-to-Sensor}
\newacronym{mpr}{MPR}{Market Penetration Rate}
\newacronym{ukf}{UKF}{Unscented Kalman Filter}
\newacronym{pdr}{PDR}{Packet Delivery Rate}
\newacronym{cpm}{CPM}{Collective Perception Message}
\newacronym{dcc}{DCC}{Decentralized Congestion Control}
\newacronym{vru}{VRU}{Vulnerable Road User}

\input{SDF_Header}

\begin{document}

\title{Track-to-Track Association for Collective Perception based on Stochastic Optimization}

\author{LAURA M. WOLF}
\affil{University of Göttingen, Germany} 

\author{VINCENT ALBERT WOLFF}

\affil{\,\!Leibniz University Hannover, Germany}

\author{SIMON STEUERNAGEL}
\affil{University of Göttingen, Germany} 

\author{KOLJA THORMANN }
\member{Member, IEEE}
\affil{University of Göttingen, Germany} 

\author{MARCUS BAUM}
\member{Member, IEEE}
\affil{University of Göttingen, Germany}

\receiveddate{
This work has been submitted to the IEEE for possible publication. Copyright may be transferred without notice, after which this version may no longer be accessible.\\
This work was partially funded by the Lower Saxony Ministry of Science and Culture under grant number ZN3493 within the Lower Saxony ``Vorab'' of the Volkswagen Foundation and supported by the Center for Digital Innovations (ZDIN).}

\corresp{{\itshape (Corresponding author: L.M. Wolf)}.}

\authoraddress{
Laura M. Wolf, Simon Steuernagel and Marcus Baum are with the Institute of Computer Science, University of Göttingen, Germany (e-mail: \{laura.wolf, simon.steuernagel, marcus.baum\}@cs.uni-goettingen.de).
Vincent Albert Wolff is with the Institute of Communications Technology, Leibniz University Hannover, Germany (e-mail: vincent.wolff@ikt.uni-hannover.de).
Kolja Thorman was with the Institute of Computer Science, University of Göttingen, Germany (e-mail: kolja.thormann@ieee.org).
}

\editor{Our source code will be available at \url{https://github.com/Fusion-Goettingen}.}
\supplementary{Color versions of one or more of the figures in this article are available online at \href{http://ieeexplore.ieee.org}{http://ieeexplore.ieee.org}.}

\markboth{AUTHOR ET AL.}{SHORT ARTICLE TITLE}
\maketitle

\begin{abstract}
Collective perception is a key aspect for autonomous driving in smart cities as it aims to combine the local environment models of multiple intelligent vehicles in order to overcome sensor limitations. 
A crucial part of multi-sensor fusion is track-to-track association. 
Previous works often suffer from high computational complexity or are based on heuristics.
We propose an association algorithms based on stochastic optimization, which leverages a multidimensional likelihood incorporating the number of tracks and their spatial distribution and furthermore computes several association hypotheses. 
We demonstrate the effectiveness of our approach in Monte Carlo simulations and a realistic collective perception scenario computing high-likelihood associations in ambiguous settings. 
\end{abstract}


\section{INTRODUCTION}

A crucial aspect of autonomous driving is environment perception. For this purpose,  autonomous vehicles are equipped with various sensors such as radar and lidar sensors or cameras. The data from these sensors can be used to estimate key characteristics of other objects in the vicinity of the autonomous vehicle, including other vehicles and vulnerable road users such as pedestrians and bicyclists. This process is called \ac{mot}~\cite{voMultitargetTracking2015}.
The estimated characteristics can include position, velocity and heading of the object, and even the object's extent~\cite{granstromTutorialMultipleExtended2022}.

An autonomous vehicle relies on its sensors, which have  a limited field of view, and suffer from occlusions due to other vehicles, buildings or the mounting point of the sensor. 
One approach to overcome these challenges is to combine the local environment model of multiple autonomous vehicles to a global model, which is called collective perception~\cite{chtourouCollectivePerceptionService2021}. In general, autonomous vehicles could  communicate with their surroundings, e.g., with each other (V2V) or with infrastructure (V2I) such as a \ac{rsu}, summed up as V2X communication.
Utilizing V2X communication the local perception of autonomous vehicles can be fused to a collective perception and then made available to all autonomous vehicles in the area in order to increase tracking accuracy and overcome occlusions~\cite{guntherPotentialCollectivePerception2015,wolffEnhancingVulnerableRoad2023}.

There are  two general approaches to multi-sensor fusion, i.e, low-level fusion and high-level fusion. In low-level fusion sensor readings are combined first and then \ac{mot} is performed on the fused measurements. 
However, transmitting all measurements can require a lot of bandwidth and the system has to be adapted to accommodate different sensor data. Furthermore, detailed sensor information might be necessary to fuse the results which most likely are not available in a collective perception scenario.
In high-level fusion each sensor or in this case autonomous vehicle tracks the objects in its vicinity using its own sensor data. Then, the estimated tracks are combined to a fused track. High level fusion requires less bandwidth as the track already consists of condensed information as opposed to all measurements, but it is therefore less accurate. Another advantage is that high-level fusion can  easily cope  with data from different types of sensors. 

Multi-sensor fusion can be carried out decentralized or in a central fusion center~\cite{chee-yeechongArchitecturesAlgorithmsTrack2000,aeberhardTracktoTrackFusionAsynchronous2012}. We focus on central fusion with a fusion center located, e.g., in an \ac{rsu}~\cite{alligDynamicDisseminationMethod2019, chtourouCollectivePerceptionService2021}. This avoids redundant computation in the autonomous vehicles and in addition an \ac{rsu} could be equipped with more computational capacity than the autonomous vehicles, which also have to perform local tracking. In the fusion center the tracks and corresponding uncertainties are fused together, which can be done using, e.g., covariance intersection~\cite{frankenImprovedFastCovariance2005}.

Before the tracks can be fused, they have to be associated (\ac{t2ta}), i.e., it has to be determined which tracks likely originated from the same object. Finding the optimal association of multiple tracks from multiple sensors given a cost function can be formulated as a multi-dimensional assignment problem, which is NP-hard for three or more sensors~\cite{gareyComputersIntractabilityGuide1990}. This problem arises in memory-less \ac{t2ta}, i.e., when all tracks are associated and fused every time step and there is no continuous global track but also when multiple objects enter the scenario at the same time and multiple new fused tracks have to be created. 
In~\cite{bar-shalomHierarchicalTrackingReal2006} it was shown that multi-sensor fusion without a global track performs better than fusion with memory and performs close to optimal. A multi-dimensional assignment problem also occurs when measurements from more than two sensors should be fused or in track birth with more than two sensors. 

\begin{figure*}[t]
    \centering
    \includegraphics[width=0.95\linewidth]{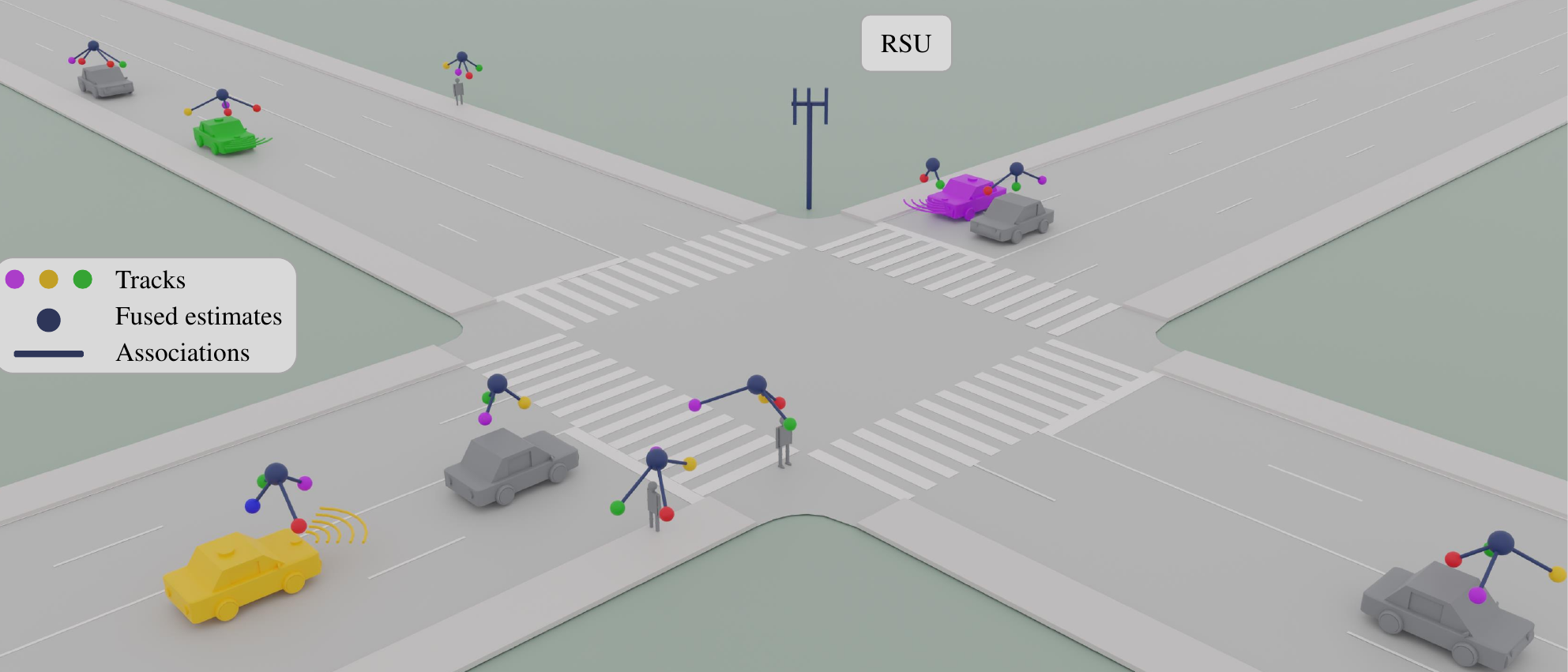}
    \caption{Visualization of the collective perception scenario with an \ac{rsu}. Pedestrians and vehicles on an intersection are tracked by intelligent vehicles (colored). Colored spheres indicate the tracks generated by the vehicle in the corresponding color. Dark blue connections represent the associations and the dark blue spheres the fused estimates.}
    \label{fig:artery_example}
\end{figure*}

\subsection{Related Works}

As the multi-dimensional assignment problem is NP-hard most algorithms reduce the computational complexity by approximating the optimization problem or applying greedy heuristics and clustering approaches.

One way to approximate the multi-dimensional assignment problem is to apply Lagrangian relaxation~\cite{debGeneralizedSDAssignment1997}. 
The constraints of the original high-dimensional problem are successively relaxed until a two-dimensional problem can be optimally solved using, e.g., the Hungarian algorithm~\cite{kuhnHungarianMethodAssignment1955}. Afterwards, the constraints are enforced again. This process is repeated until the algorithm converges. This approach has been extensively studied, however, a drawback is that a full cost matrix has to be computed beforehand, which grows exponentially in the number of sensors.

Other optimization-based approaches addressing the multi-dimensional assignment problem apply linear programming as a relaxation to the integer problem and then reconstruct an integer solution. This, however, has not yet been applied to \ac{t2ta} but has been applied to \ac{mot} and multi-frame association~\cite{stormsLPbasedAlgorithmData2003}, where data from multiple frames from the same sensor are associated. 

Another approach computes a permutation matrix to associate the tracks of pairwise sensors using deterministic annealing~\cite{leeMultitargetTracktotrackFusion2018}. It is not easily applicable to scenarios with many sensors as ambiguities in the pairwise associations are handled heuristically. 

The GRASP approach presented in~\cite{robertsonSetGreedyRandomized2001, murpheyGreedyRandomizedAdaptive1997} combines a greedy approach with local optimization. First, it selects one of the highest scoring tuples at random, eliminates all conflicting tuples and repeats the process until all tracks are associated. Then, this association is improved by local exchanges. However, the computation of all possible or likely tuple can be computational infeasible.

The approach presented in~\cite{houenouTracktotrackAssociationMethod2012} greedily selects pairwise associations of two tracks up to a maximum distance, as long as the resulting cluster does not violate the sensor constraint. However, if both tracks are already part of a cluster the pair is skipped. This leads to suboptimal results where one object is represented by several small clusters in scenarios with more than three sensors. 
We addressed this issue in our previous work~\cite{wolfTracktotrackAssociationBased2023} by adding a merging step and could greatly improve the performance in scenarios with many sensors. This greedy variant with merging is very similar to single-linkage agglomerative clustering, with the additional sensor constraint.

The clustering approach presented in~\cite{liMultisourceHomogeneousData2017} first uses density-based clustering ignoring the sensor constraint and then uses a heuristic to split oversized clusters. If the sensor requirement needs to be enforced, conflicting tracks are removed from the clusters. This could lead to many singleton tracks and suboptimal results in \ac{t2ta} scenarios with no clutter.

In~\cite{chummunFastDataAssociation2001} clustering was applied as a first step to reduce the size of the association problem. Possible clusters are determined via gating and enumeration forming an association tree. Then, the tracks belonging to an oversized or conflicting cluster are associated using Lagrangian relaxation~\cite{debGeneralizedSDAssignment1997}.
 
\subsection{Contribution}

In this paper we 
\begin{itemize}
    \item propose a novel \ac{t2ta} algorithm based on \ac{so},
    \item propose a cluster likelihood that also incorporates the uncertainty of the cluster center, and
    \item evaluate the proposed method as well as relevant state of the art in Monte Carlo simulations and a complex collective perception scenario that investigates different communication strategies.
\end{itemize}
The proposed algorithm is computationally efficient with a quadratic complexity in the number of tracks and does not rely on distance thresholds that have to be adapted to each scenario.

This article is an extended version of our conference article~\cite{wolfTracktotrackAssociationBased2023}, where the \ac{t2ta} algorithm was first proposed. However, in~\cite{wolfTracktotrackAssociationBased2023}, the algorithm was solely evaluated in Monte Carlo simulations. In this article, we complement the experiments with a realistic collective perception scenario including different communication strategies. A deterministic variant of our \ac{so} approach was published in~\cite{wolfTracktotrackAssociationBased2024}.

\subsection{Structure}
This paper is structured as follows: First, we formally describe the association problem in Sec.~\ref{sec:problem} and  derive the likelihood that is used to estimate how likely a set of tracks stems from the same origin in Sec.~\ref{sec:likelihood}. Then, we propose the novel association approach based on \ac{so} and briefly revisit greedy association methods and Lagrangian relaxation for SD-assignment in Sec. \ref{sec:SO} and \ref{sec:sota}, respectively. In Sec.~\ref{sec:simulation_setup} we describe the simulation setup for the Monte Carlo simulation and the details of the collective perception scenario. The results are presented and discussed in Sec.~\ref{sec:results} and Sec.~\ref{sec:discussion} and the paper is concluded in Sec.~\ref{sec:conclusion}.

\section{PROBLEM DESCRIPTION}
\label{sec:problem}

In this paper we consider the problem of multi-sensor fusion in a fusion center without a global track.

Each timestep we obtain a set of tracks  ${\trackset=\{1,\dots, \numtracks\}}$, where the tracks are arbitrarily numbered from 1 to $\numtracks$. The tracks consist of corresponding states $\state{1},\dots, \state{\numtracks}$ and covariances $\cov{1},\dots,\cov{\numtracks}$. As we only consider one timestep the time index is omitted. 
The objective is to group the tracks into clusters such that all tracks in one cluster originate from the same object.

The tracks stem from a set of sensors ${\sensorset = \{1,\dots,\numsensors\}}$, located at positions $\sensorpos{1},\dots,\sensorpos{\numsensors}$. 
The function $s(t)$ returns the sensor that generated track $t\in\trackset$, whereas $S(\cluster{c})=\{s(t)\mid t\in\cluster{c}\}$ returns the sensors that generated the tracks in a set of tracks, also called a cluster $\cluster{c}\subseteq \trackset$. 
It is assumed that each sensor only transmits at most one track per object. 
Hence, in a valid association, all tracks in a cluster must originate from different sensors, i.e., $|S(\cluster{c}) |= |\cluster{c}|$.
Fig.~\ref{fig:artery_example} shows an exemplary time step from the collective perception scenario, where several objects (cars and pedestrians) are tracked by three cars equipped with sensors yielding tracks in the respective color. Dark blue connections and spheres indicate the correct association and the fused estimate. 

Given a cost function $\mat{d}$ representing the likelihood of a set of tracks stemming from the same object, the multi-sensor association problem can be formulated as a multi-dimensional assignment problem:
\begin{align}
    \min_{\rvec{\rho}_{i_{1}i_{2}\dots i_{\numsensors}}} 
    \sum_{i_1=0}^{|\trackset_1|} \sum_{i_2=0}^{|\trackset_2|}\dots \sum_{i_\numsensors=0}^{|\trackset_\numsensors|} \mat{d}_{i_{1}i_{2}\dots i_{\numsensors}}\rvec{\rho}_{i_{1}i_{2}\dots i_{\numsensors}}
\end{align}
such that 
\begin{align}
\begin{split}
    \sum_{i_2=0}^{|\trackset_2|} \dots \sum_{i_\numsensors=0}^{|\trackset_\numsensors|} \rvec{\rho}_{i_{1}i_{2}\dots i_{\numsensors}} =1, & \quad i_1=1, \dots, |\trackset_1|\\
    \sum_{i_1=0}^{|\trackset_1|} \dots \sum_{i_\numsensors=0}^{|\trackset_\numsensors|} \rvec{\rho}_{i_{1}i_{2}\dots i_{\numsensors}} =1, & \quad i_2=1, \dots, |\trackset_2|\\
    \vdots\quad\quad\quad\quad\quad\quad&\quad \quad\quad \vdots \\
    \sum_{i_1=0}^{|\trackset_1|} \dots \sum_{i_{\numsensors-1}=0}^{|\trackset_{\numsensors-1}|} \rvec{\rho}_{i_{1}i_{2}\dots i_{\numsensors}} =1, & \quad i_{\numsensors}=1, \dots, |\trackset_\numsensors|  \label{eq:mda_constraints}
\end{split}
\end{align}
Here, $\trackset_s$ contains all tracks originating in sensor $s$ numbered from 1 to $|\trackset_s|$. $\rvec{\rho}_{i_{1}i_{2}\dots i_{\numsensors}}$ is a binary value indicating if tracks $i_{1}i_{2}\dots i_{\numsensors}$ are associated together with corresponding cost $\mat{d}_{i_{1}i_{2}\dots i_{\numsensors}}$. The value 0 for a track indicates that no track from this sensor is associated in this cluster. The constraints \eqref{eq:mda_constraints} ensure that each track (except for 0) is associated to exactly one cluster. The objective is to minimize the cost summed over all possible combinations of tracks. 

Maintaining an association and a cost variable for each possible cluster is infeasible even for a moderate amount of sensors and objects. 
We therefore represent an association of the tracks with a joint association variable~$\jointasso$, that maps each track to a cluster: 
\begin{equation}
\jointasso: \trackset \mapsto \mathbb{N}^{\numtracks}, \jointasso = [\theta_1,\dots,\theta_{\numtracks}] \enspace,
\label{eq:jointasso}
\end{equation}
where track $t$ is mapped to cluster $\cluster{\theta_t}$.
The clusters are arbitrarily numbered and $\clusterlist = \{c \mid \theta_t = c, t\in\trackset\}$ maintains the current cluster indices. The cluster $\cluster{c} = \{t \mid \theta_t = c\}$ contains all tracks that are mapped to cluster $c$, such that for all $t\in\trackset$ and $c\in\clusterlist$ we have $t\in \cluster{c} \iff \theta_t = c$.

The joint association $\jointasso$ is ambiguous, e.g., ${\jointassoit{1}=[1,2,1]}$ resembles the same association as ${\jointassoit{2} = [3,2,3]}$. However, an unambiguous representation can be accomplished if the order of the tracks is fixed. Numbering the clusters consecutively in the order of first appearance in $\jointasso$ yields an unambiguous association to, e.g., compare two associations.


\section{PROPOSED LIKELIHOOD}
\label{sec:likelihood}
In order to determine which association is more plausible than another we employ a likelihood function.

The likelihood of a joint association ${p(\state{1},\dots,\state{\numtracks}\mid \jointasso)}$ can be decomposed into the likelihoods of each cluster $l(\cluster{c}) $ as tracks associated to different objects are independent~\cite{kaplanAssignmentCostsMultiple2008}
\begin{equation}
\begin{split}
p\left(\state{1},\dots,\state{\numtracks}\mid \jointasso\right) =& \prod_{c\in\clusterlist} p\left(\left\{\state{t'}\mid t'\in\cluster{c}\right\} \mid\cluster{c}\right) \\
=& \prod_{c\in\clusterlist} l({\cluster{c}}) \enspace.    
\end{split}
\end{equation}

The cluster likelihood $l(\cluster{c})$ is then comprised of a likelihood for the cluster size $l_c(\cluster{c})$ and a spatial likelihood $l_s(\cluster{c})$
\begin{equation}
    l(\cluster{c})=l_c(\cluster{c})\cdot l_s(\cluster{c})\enspace.
\end{equation}
It is assumed that each sensor has perceived an object with probability $p_D(\mat{x}, s)$, which depends on the state of the object $\mat{x}$ and the sensor $s$.

The object's state used for the detection probability is estimated as the track fusion of the tracks in the cluster:
\begin{subequations}
\begin{align}
    \clustercenter & = \clustercov  \sum_{t \in \cluster{c}} \cov{t}^{-1} \state{t} \enspace , \\
    \clustercov & =  \left( \sum_{t \in \cluster{c}} \cov{t}^{-1} \right)^{-1} \enspace ,
\end{align}
\end{subequations}
where it is assumed that the track errors are independent. 
Hence, the likelihood of the cluster size can be computed as:
\begin{equation}
    l_c(\cluster{c})=\prod_{s\in S(\cluster{c})} p_D(\clustercenter, s) \cdot  \prod_{s\in \sensorset \setminus S(\cluster{c})} 1 - p_D(\clustercenter, s)
    \enspace.
\end{equation}

Given that the track errors are independent, the spatial likelihood given the true state $\mat{x}$ can be written as
\begin{equation}
     l_s(\cluster{c}) = \prod_{t\in\cluster{c}} \mathcal{N}\left(\state{t};\mat{x},  \cov{t} \right) \enspace.
\end{equation}
Approximating the true state $\mat{x}$ as a Gaussian density with mean $\clustercenter$ and covariance $\clustercov$, we get
\begin{subequations}
\label{eq:likelihood}
\begin{align}
    l_s(\cluster{c}) &\approx \prod_{t\in\cluster{c}} \int_\mat{x} \mathcal{N}\left(\state{t};\mat{x},  \cov{t} \right) \cdot \mathcal{N}\left(\mat{x}; \clustercenter, \clustercov \right) \mathrm{d} \mat{x}\\
    &= \prod_{t\in\cluster{c}} \mathcal{N}\left(\state{t};\clustercenter, \clustercov + \cov{t} \right) \enspace.
\end{align}
\end{subequations}

In contrast to the generalized likelihood introduced in\cite{kaplanAssignmentCostsMultiple2008}, our proposed likelihood includes the uncertainty of the estimated cluster center, which is especially relevant for small clusters. In the case that all tracks have the same covariance \eqref{eq:likelihood} simplifies to~\cite[Eq. (2)]{wolfTracktotrackAssociationBased2023}.

\section{STOCHASTIC OPTIMIZATION}
\label{sec:SO}
The \ac{t2ta} approach that we propose is based on \ac{so} and optimizes the likelihood by randomly making alterations to the current association. It starts with a valid association, e.g., that each track is its own cluster and then iteratively samples a new action for every track. We consider the following actions for the current track:
\begin{itemize}
    \item $r$: remain in current cluster, the association does not change,
    \item $s$: split current track from cluster and create a singleton, 
    \item $m_c$: move current track to $\cluster{c}$,
    \item $M_c$: merge current cluster with $\cluster{c}$.
\end{itemize}
For all actions 
\begin{equation}
    \mat{A}=[r,s,m_1,\dots,m_{|\clusterlist|}, M_1,\dots,M_{|\clusterlist|}] \enspace,
    \label{eq:actions}
\end{equation}
we then have the corresponding likelihoods
\begin{equation}
    \mat{p_A} \propto[p^r, p^s,p^{m_1},\dots,p^{m_{|\clusterlist|}}, p^{M_1},\dots,p^{M_{|\clusterlist|}}]\enspace.    
    \label{eq:sample_lik}
\end{equation}
The likelihood $p^a$ of action $a$ is given by the ratio of likelihoods 
\begin{equation}
p^a=\frac{p\left(\state{1},\dots,\state{\numtracks}\mid \jointassoit{a}\right)}{p\left(\state{1},\dots,\state{\numtracks}\mid \jointasso\right)}\enspace,    
\end{equation}
where $\jointasso$ is the current joint association and $\jointassoit{a}$ the resulting joint association after performing action $a \in \mat{A}$.

As the association does not change, when the current track remains in its cluster, we have 
\begin{equation}
    p^r = 1\enspace.
\end{equation}

Since we consider the likelihood ratio, the likelihoods of clusters that are not affected by the action cancel out. This yields the following likelihood for creating a singleton cluster from the current track:
\begin{equation}
    p^s = \begin{cases} 
        \displaystyle\frac{l(\{t\}) \cdot l(\cluster{\theta_t}\setminus\{t\})}{l(\cluster{\theta_t})} \enspace & \text{if }|\cluster{\theta_t}|>1\enspace, \\
        0\enspace &  \text{else} \enspace.
    \end{cases}
    \label{eq:prob_singleton}
\end{equation}
If the track is already a singleton cluster, the likelihood of this action is zero, to avoid double counting. 

For the same reason we do not consider moving the track to its current cluster. Furthermore, it is only possible to move the current track to another cluster if all tracks originate from different sensors, i.e., $s(t)\notin S(\cluster{c})$:
\begin{equation}
    p^{m_c} = \begin{cases}
        \displaystyle\frac{ l(\cluster{c} \cup \{t\})\cdot l(\cluster{\theta_t}\setminus\{t\})}{l(\cluster{c})\cdot l(\cluster{\theta_t})}  
        & \text{if } s(t)\notin S(\cluster{c}) \enspace,\\
        0 & \text{else} \enspace.
    \end{cases}
    \label{eq:prob_move}
\end{equation}
In the same way the likelihood for merging the current cluster with another cluster is given, as long as the tracks all stem from different sensors:
\begin{equation}
    p^{M_c} = \begin{cases}
    \displaystyle \frac{l(\cluster{c} \cup \cluster{\theta_t})}{l(\cluster{c})\cdot l(\cluster{\theta_t})} &
    \text{if } S(\cluster{c}) \cap S(\cluster{\theta_t}) = \emptyset\enspace,\\
    0 & \text{else} \enspace.
    \end{cases}
    \label{eq:prob_merge}
\end{equation}

It is possible to add a gating step in order to decrease computation time, thus, only considering clusters that are within a certain gating distance $d_g$
 \begin{equation}
    \left\Vert \overline{\mat{x}}^c - \mat{x}_t\right\Vert \leq d_g\enspace.
 \end{equation} 
 
The full algorithm can be found in Fig.~\ref{alg:random_so}. The algorithm performs $N$ sweeps over all tracks and samples and performs an action for every track. After every action the current sample is saved. The initial association assigns every track to its own cluster. To facilitate sampling we set the maximum applied detection probability to 0.97 as otherwise all clusters that do not contain tracks from all sensors have zero likelihood.

The main part of the \ac{so} algorithm regarding computational complexity is the computation of $\mat{p_A}$, which has a complexity of $\mathcal{O}(\numtracks)$ as all tracks have to be considered once to compute the likelihoods and there are at most $\numtracks$ clusters. In total the algorithm has a complexity of $\mathcal{O}\left(N\cdot \numtracks^2\right)$ to compute $N\cdot\numtracks$ joint associations.

\begin{figure}[t]
\begin{algorithmic}[1]
    \Require $\trackset$, $\sensorset$, $N$, $p_D$
    \Ensure $\jointassoit{1,1} \dots , \jointassoit{N,\numtracks}$
    \State $\jointasso \gets [1, \dots, \numtracks]$
    \State $k\gets\numtracks + 1$
    \For{$n=1,\dots,N$}
        \For{$t=1,\dots,\numtracks$}
            \State calculate  $\mat{A}$ and $\mat{p_A}$ according to \eqref{eq:actions} and \eqref{eq:sample_lik}
            \State normalize $\mat{p_A}$
            \State $a \gets \randchoice(\mat{A},\mat{p_A})$
            \If{$a=s$}
                \State $\theta_t \gets k$
                \State $k \gets k+1$
            \ElsIf{$a = m_c, c\in \clusterlist$}
                \State $\theta_t \gets c$
            \ElsIf{$a = M_c, c\in \clusterlist$}
                \State $\theta_{t'} \gets c, t'\in \cluster{\theta_t}$
            \EndIf
            \State $\jointassoit{n,t} \gets \jointasso$
        \EndFor
    \EndFor
\end{algorithmic}
\caption{Stochastic optimization algorithm for \ac{t2ta}}
\label{alg:random_so}
\end{figure}

\section{COMPARISON METHODS}
\label{sec:sota}
For comparison we also consider the following methods: the greedy approach by~\cite{houenouTracktotrackAssociationMethod2012} which we extended by an additional merging step in~\cite{wolfTracktotrackAssociationBased2023}, the Lagrangian relaxation approach from~\cite{debGeneralizedSDAssignment1997} and a simple heuristic, where we iteratively apply optimal two-dimensional assignments sensor-wise. The methods are detailed in the following subsections. 

\subsection{Greedy with merging}
\label{sec:greedy}

The greedy approach was originally proposed in~\cite{houenouTracktotrackAssociationMethod2012} and we extended it with an additional merging step in~\cite{wolfTracktotrackAssociationBased2023} as the original approach did not necessarily yield satisfactory results for scenarios with more than three sensors. 
The approach works with pairwise distances between tracks and the original paper proposed a distance based on the track history, where the distances between two objects over the last time steps are averaged. In~\cite{wolfTracktotrackAssociationBased2023} we simply applied the Euclidean distance, as we only considered single time steps. In this work we apply the negative log cluster likelihood as proposed in Sec.~\ref{sec:likelihood} on all pairs of tracks. Therefore, the underlying distance is the same for all algorithms, although the greedy approach only considers pairwise distances and hence does not incorporate the likelihood for the final cluster cardinality. 

Tracks originating from the same sensor cannot be put in the same cluster, their distance is set to a maximal value $d_m$. The same applies for tracks that are far apart, i.e., the track distance is greater than a threshold $d_t$. Furthermore, only the lower triangular part of the distance matrix is considered, as it is symmetric. 

The algorithm first assigns all tracks to singleton clusters, i.e., $\jointasso=[1,\dots,\numtracks]$ and then considers all track pairs in ascending order of their distance.
If both tracks are singletons, they are moved to the same cluster.
If one track is a singleton, it is moved to the cluster of the other track, if it doesn't already contain a track from the same sensor.
If both tracks are in a cluster, the original algorithm would skip the pair~\cite[Step 4) c) in Algorithm 1]{houenouTracktotrackAssociationMethod2012}. This can lead to the situation that with more than three sensors, tracks belonging to the same object, are grouped in two or more clusters, that will never be merged. We proposed in~\cite{wolfTracktotrackAssociationBased2023} to merge both clusters if the sensors of all tracks are distinct, i.e., $S(\clusterlist_{\theta_i})\cap S(\clusterlist_{\theta_j}) = \emptyset$.

The full algorithm can be found in Fig.~\ref{alg:greedy}. We will refer to both variants as \emph{Greedy (merge)} and \emph{Greedy (no merge)}, respectively.
The Greedy algorithm has a complexity of $\mathcal{O}(\numtracks^2)$ as the distance matrix has the dimensions $\numtracks\times\numtracks$ and all track pairs are considered. 

\begin{figure}[t]
\begin{algorithmic}
    \Require $\trackset$
    \Ensure $\jointasso$
    \State $\jointasso \gets [1, \dots, \numtracks]\in \mathbb{N}^{\numtracks}$
    
    \For{$t_i,t_j\in \mat{T}$}
    \State $D_{i,j} \gets l_s(\{t_i,t_j\})$
    \If{$j\geq i$ or $D_{i,j}>d_t$ or $s(t_i)=s(t_j)$} \State $D_{i,j}=d_m$\EndIf
    \EndFor
    \While{$\min(D)<d_m$}
        \State $i,j \gets \argmin(D)$
        \If {$|\clusterlist_{\theta_{t_i}}|=1$ and $|\clusterlist_{\theta_{t_j}}|=1$}
            \State $\theta_{t_i}\gets\theta_{t_j}$
        \ElsIf{$|\clusterlist_{\theta_{t_i}}|=1$ and $s(t_i)\notin S(\clusterlist_{\theta_j})$}
            \State $\theta_{t_i}\gets\theta_{t_j}$
        \ElsIf{$|\clusterlist_{\theta_{t_j}}|=1$ and $s(t_j)\notin S(\clusterlist_{\theta_i})$}
            \State $\theta_{t_j}\gets\theta_{t_i}$
        \ElsIf{$\theta_{t_j}\neq\theta_{t_i}$ and $S(\clusterlist_{\theta_i})\cap S(\clusterlist_{\theta_j}) = \emptyset$}
            \State $\theta_{t}\gets\theta_{t_j}, \enspace t\in\clusterlist_{\theta_{t_i}}$
        \EndIf
        \State $D_{t_i,t} \gets d_m,\enspace t\in\{t'\in\trackset \mid s(t')= s(t_j)\} $
        \State $D_{t,t_j} \gets d_m,\enspace t\in\{t'\in\trackset \mid s(t')= s(t_i)\}$
    \EndWhile

\end{algorithmic}
\caption{Greedy (merge) algorithm~\cite{houenouTracktotrackAssociationMethod2012, wolfTracktotrackAssociationBased2023}.}
\label{alg:greedy}
\end{figure}

\subsection{Sequential optimal 2D assignments}    
The idea of this approach is to create a simple baseline association algorithm by associating all tracks sensor-wise with optimal two-dimensional assignments that can be achieved using, e.g, the Hungarian algorithm~\cite{kuhnHungarianMethodAssignment1955}. 

The tracks from the first sensor are all added to their own clusters. Then for each sensor its tracks are associated to the tracks that have last been added to a cluster. The negative log likelihood of the two tracks is used as the distance as in Sec.~\ref{sec:greedy}. The tracks are added to the clusters they have been assigned to. However, if two tracks are assigned that exceed the threshold distance $d_t$ the track is added to a new cluster. 

As the \emph{sensor-wise optimization} performs $\numsensors-1$ optimal 2D assignments the total complexity of the algorithm is $\mathcal{O}((\numsensors-1)\cdot \numobjects^3)$, where $\numobjects$ is the number of objects.

\subsection{Lagrangian Relaxation}

Finally, we consider the SD-assignment approach based on Lagrangian relaxation approach from~\cite{debGeneralizedSDAssignment1997}. The approach optimizes a multidimensional assignment problem based on a cost matrix. As the direct optimization is infeasible, in each optimization step the constraints are successively relaxed until a two-dimensional assignment problem can be optimally solved using the Hungarian or Auction algorithm. Then, the constraints are enforced again and the process is repeated. 

The cost matrix contains a cost for each possible cluster. As likelihoods of clusters of different size cannot be compared due to different dimensions, likelihood ratios of cluster likelihoods are considered, where the cluster likelihood is divided by the likelihood that all tracks in the cluster are  singletons. Then the negative log likelihood ratio is used as cost for the SD-assignment approach. 

The approach has a computational complexity of $\mathcal{O}(\numsensors - 1)\cdot \numobjects^3$ like the sensorwise optimization, although with a higher constant factor, as the optimization is iterated several times. In addition, the cost matrix has to be computed, which has a space complexity of $\mathcal{\numobjects^\numsensors}$.

We use the MATLAB implementation\footnote{\url{https://ch.mathworks.com/help/fusion/ref/assignsd.html}} in the Sensor Fusion and Tracking Toolbox for our experiments and refer to this method as \emph{SD-assign}.

\section{SIMULATION SETUP}
\label{sec:simulation_setup}

To demonstrate the effectiveness of our approach we conducted two experiments, a Monte Carlo simulation and a collective perception scenario. The collective perception scenario is simulated using SUMO~\cite{lopezMicroscopicTrafficSimulation2018} which creates realistic trajectories coupled with the Artery simulator~\cite{rieblArteryExtendingVeins2015} which provides the simulation of realistic V2X communication according to ETSI ITS-G5 \cite{ITS2020}. 

\subsection{Monte Carlo Simulation}

First, we evaluated the association algorithms on two Monte Carlo simulations with static objects. The small scenario consists of 8 objects and 5 sensors in a surveillance area of $\SI{30}{\m} \times \SI{30}{\m}$. The big scenario covers $\SI{50}{\m} \times \SI{50}{\m}$ with 20 objects and 12 sensors.
The objects are distributed uniformly at random within the surveillance area. The sensors generate a track for each object with probability $p_D$, which is the same for all sensors. The track is generated by adding Gaussian white noise with covariance $\cov{} = \sigma^2\mat{I}_2$. Hence, all tracks have the same covariance $\cov{}$. We considered two noise scenarios: low noise ($\sigma=1$) and high noise ($\sigma=2$). 

\begin{figure}
    \centering
    \subfloat[$\sigma=1.0$.\label{fig:mc_sample_low_noise}]{\includegraphics[width=0.45\linewidth]{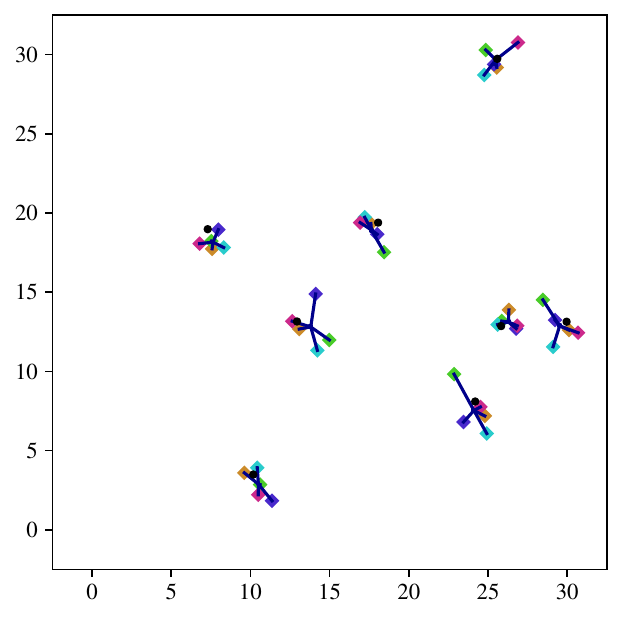}}
    \qquad
    \subfloat[$\sigma=2.0$.\label{fig:mc_sample_high_noise}]{\includegraphics[width=0.45\linewidth]{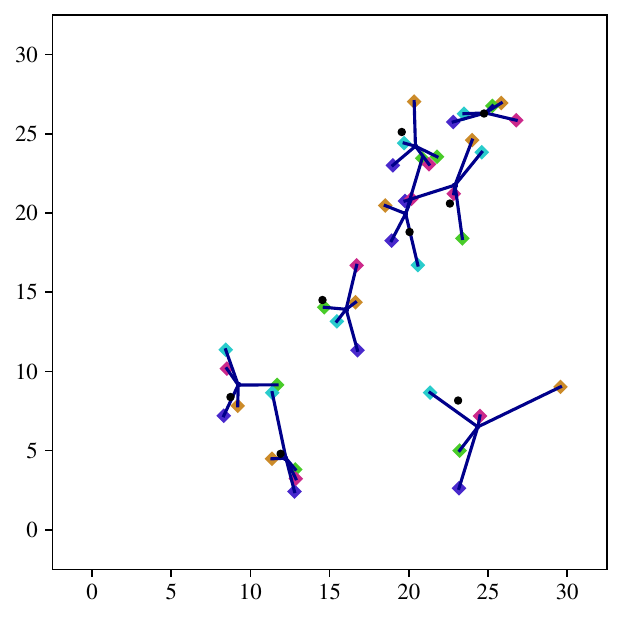}}
    \caption{Sample of small Monte Carlo simulation with detection probability $p_D=1.0$ and ground truth association.}
    \label{fig:mc_sample}
\end{figure}

An example of the small Monte Carlo simulation can be found in Fig.~\ref{fig:mc_sample}.

\subsection{Collective Perception Scenario}

The second simulation aims to emulate a collective perception scenario, where autonomous vehicles communicate with infrastructure modules to combine their local environment perception. We made use of the Artery simulator which couples traffic simulation and network simulation to simulate V2X communication. Our scenario focuses on an urban intersection with an \ac{rsu}. A certain percentage of the vehicles, defined by the \ac{mpr} is equipped with sensors and communicates with the \ac{rsu}. Each intelligent vehicle tracks objects in its vicinity, i.e., other vehicles and pedestrians. These tracks are then communicated to the \ac{rsu}, where they are associated and fused. The following subsections describe the details of each step.

\subsubsection{Local tracking}
Each intelligent vehicle is equipped with a camera sensor, which has a field of view of $\SI{360}{\degree}$ and a range of \SI{85}{\m}. Artery returns the ground truth positions of objects in the sensor range, that are not occluded by other vehicles or buildings. 
We generated point measurements by adding white Gaussian noise with covariance $\mat{R}= 2^2\mat{I}_2$. To facilitate the local tracking, we did not consider object-to-measurement data association, i.e., we assumed that it is known which measurement belongs to which object. 
The simulator generates detections every \SI{0.1}{\s}. Furthermore, The track is invalidated if the object is not detected for \SI{0.4}{\s}. Then, a new track is created, when the object is detected again.

We implemented an \ac{ukf}~\cite{wanUnscentedKalmanFilter2000} and modeled the state as 
\begin{equation}
    \state{} = [x_0, x_1, \dot{x}_0, \dot{x}_1, \omega]^T \enspace,
\end{equation}
with position $[x_0,x_1]^T$, Cartesian velocity $[\dot{x}_0, \dot{x}_1]^T$ and yaw rate $\omega$. We applied the coordinated turn model~\cite{rothEKFUKFManeuvering2014} $\state{k+1} = f(\state{k}) + \mat{G}\mat{w}_k$ with $f(\state{k})$ and $\mat{G}$ according to~\cite[Eqs. (6), (7b)]{rothEKFUKFManeuvering2014} and a nearly constant velocity model, when the yaw rate is close to zero. The process noise is set to $\mat{w}_k\sim \mathcal{N}(\mat{0}, \diag{(5^2, 5^2, (0.08\pi)^2)}$. Measurements are generated according to $\mat{z}_k = \mat{H}\state{k}+\mat{v}_k$ with $\mat{H}=\begin{bmatrix}1&0&0&0&0\\0&1&0&0&0\end{bmatrix}$ and $\mat{v}_k\sim \mathcal{N}(\mat{0}, \mat{R})$.

The initial position and velocity are derived from the first two measurements, the initial yaw rate is set to zero.
\subsubsection{Communication}
In our simulation, \acp{cpm} are transmitted using the ITS-G5 protocol stack within the Artery framework, operating over a 10 MHz channel in the 5.9 GHz band. \ac{dcc} is disabled to restrict message generation based on the generation rules, which will be explained further in the next paragraphs.
Each intelligent vehicle sends a \ac{cpm} at a minimum interval of \SI{0.1}{\s}. The \ac{cpm} includes information about the sender, e.g., the sender's ID, current position, and field of view. Furthermore, it contains all submitted tracks, including state, covariance, time of measurement, and track ID, which is only locally unique.

We consider two sets of rules for generating \acp{cpm}: full communication and the ETSI dynamic rules. In the full communication scenario, all current tracks, i.e., those that have been updated within the last \SI{0.1}{\s}, are included in the \ac{cpm}.
The ETSI message generation rules~\cite{europeantelecommunicationsstandardsinstituteetsiETSITS1032023} are designed to reduce communication load by avoiding the transmission of tracks with marginal changes. These rules differentiate between tracks of \acp{vru} (e.g., pedestrians or bicyclists) and other vehicles. Newly generated tracks are always included. \ac{vru} tracks are included if they have not been sent for more than \SI{0.5}{\s}. If one \ac{vru} track is included, all known \ac{vru} tracks are also included. Vehicle tracks are included if they have not been sent for more than \SI{1}{\s} or if the object's state has changed significantly. A track is considered significantly changed if, since its last transmission, its position has shifted by more than \SI{4}{\m}, its speed has changed by more than \SI{0.5}{\m\per\s}, or its orientation has altered by more than \SI{4}{\degree}.

In the \ac{rsu}, the received tracks are accumulated over \SI{0.1}{\s} for the full communication strategy and over \SI{1}{\s} for the ETSI communication rules. They are propagated to the current time to account for possible delays. Additionally, tracks that have not been updated for more than \SI{1}{\s} are removed. Finally, the tracks are associated every \SI{0.1}{\s}.

\subsubsection{Adaptation of the detection probability} 
We consider two possible strategies to model the detection probability in the collective perception scenario: constant overall and depending on the sensor position. 

The overall constant detection probability is estimated as: 
\begin{equation}
    p_D(\bar{\mat{x}}, s) = p_D = \rho \cdot\frac{\numtracks}{\numsensors\cdot \max_{s'\in\sensorset}|\{t\in\trackset| s(t) = s'\}|}\enspace.
\end{equation}
The maximum number of tracks generated by one sensor is $\max_{s'\in\sensorset}|\{t\in\trackset| s(t) = s'\}|$ and $\rho$ estimates the fraction of objects that can at most be seen by one sensor in the surveillance area. We use $\rho=\frac{1}{4}$ .

The detection probability depending on the sensor position is constant and high within the field of view of the sensor (including a margin of $\SI{10}{\metre}$ to account for movement and propagation errors) of the sensor and low otherwise:
\begin{equation}
    p_D(\overline{\mat{x}}, s)  = 
        \begin{dcases}
        0.97, & || \overline{\mat{x}} - \mat{s}_s || < \SI{95}{\m}\enspace,\\
        0.15, & \text{else\enspace.}
        \end{dcases}
\end{equation}

\subsubsection{Scenarios}
For each communication scenario we consider three market penetration rates, i.e., the percentage of the vehicles that are equipped with sensors. The resulting number of objects and sensors can be found in Table~\ref{tab:artery_stats}. The association is performed after \SI{15}{\s}, when most of the objects have appeared in the scenario, for \SI{30}{\s}, i.e., 300 association steps. Fig.~\ref{fig:artery_example} shows the central intersection of the collective perception scenario with MPR $=\SI{50}{\percent}$. 

\begin{table}
    \centering
    \caption{Number of objects and sensors for the full communication and ETSI collective perception scenarios and three market penetration rates.}
    \label{tab:artery_stats}
\begin{tabular}{lllrrr}
\toprule
 &  & MPR & min & max & mean \\
\midrule
 & Objects &  & 100 & 148 & 123.8 \\
\midrule
\multirow[t]{3}{*}{Full Comm.} & Sensors & 25 & 6 & 30 & 15.3 \\
 & Sensors & 50 & 13 & 41 & 27.6 \\
 & Sensors & 100 & 2 & 63 & 33.4 \\
\midrule
\multirow[t]{3}{*}{ETSI} & Sensors & 25 & 7 & 30 & 15.5 \\
 & Sensors & 50 & 18 & 46 & 29.8 \\
 & Sensors & 100 & 34 & 83 & 58.8 \\
\bottomrule
\end{tabular}

\end{table}

In Table~\ref{tab:pdr} We show the \ac{pdr}, a key metric for communication reliability, for both communication scenarios. The \ac{pdr} is determined on a per-vehicle basis. For each individual transmitting vehicle, its \ac{pdr} is calculated as the ratio of the number of packets it sent that were successfully received by other nodes to the total number of packets it transmitted during the analysis window. The total number of transmitted packets is inferred from the span of the packet sequence numbers observed at the receivers.

\begin{table}
    \centering
    \caption{Mean \ac{pdr} for both communication scenarios and three different market penetration rates.}
    \label{tab:pdr}

\begin{tabular}{lrrr}
\toprule
 & \multicolumn{3}{c}{PDR} \\
MPR & 25 & 50 & 100 \\
\midrule
Full Comm. & 1.00 & 0.91 & 0.60 \\
ETSI & 1.00 & 0.99 & 0.77 \\
\bottomrule
\end{tabular}

\end{table}

\subsection{Evaluation}

For the Monte Carlo simulation we evaluated the association algorithms on the small and big scenario with varying detection probabilities. We applied \ac{so} with $N=100$ sweeps in the small scenario and $N=200$ sweeps in the big scenario and a gating threshold of $d_g=6\cdot\sigma$. In the collective perception scenario we applied $N=50$ and $d_g=\SI{15}{\m}$. Furthermore in the collective perception scenario we evaluated \ac{so} with different strategies to estimate the correct detection probability, i.e., \ac{so-c} and \ac{so-ds}. 

For both Greedy variants and the sensor-wise optimization we applied a maximum distance of $d_t = 15$ and $d_t = 20$ for the Monte Carlo scenario and the collective perception scenario respectively. 

The SD-assign was performed with maximum 300 steps and a desired gap of 0.001. As the computation of the cost matrix for SD-assign is intractable for the bigger scenarios, we only evaluated it on the small Monte Carlo scenario. 

To evaluate the association, the tracks in each cluster are fused using fast covariance intersection~\cite{frankenImprovedFastCovariance2005}. As a metric the \ac{gospa}~\cite{rahmathullahGeneralizedOptimalSubpattern2017} between the position of the fused estimate and the ground truth position is computed. The \ac{gospa} considers the error of merging two clusters as missed detections and the error of creating too many small clusters as false detections and, furthermore, the accuracy of the resulting fused estimate as localization error. We applied the cutoff distance $c=\SI{10}{\m}$ and $p=1$.

To analyze the convergence behavior of \ac{so} we evaluated the relative \ac{gospa} error, i.e. the \ac{gospa} error divided by the error of the ground truth association, for the big scenario with $\sigma=2.0$ and $p_D=0.8$. 

To assess our likelihood we evaluated one exemplary Monte Carlo scenario and one collective perception scenario with alternative objective functions, namely the generalized likelihood from~\cite{kaplanAssignmentCostsMultiple2008} and the Euclidean distance for both Greedy variants and the sensor-wise optimization. For the generalized likelihood we applied the same distance threshold as for our likelihood and for the Euclidean distance we applied $d_t=\SI{10}{\m}$.

\section{RESULTS}
\label{sec:results}

In this section we first present the results of the Monte Carlo simulation regarding the GOSPA metric and the convergence behavior of SO. Then, we present the GOSPA results of the collective perception scenario as well as the exemplary likelihood comparisons.

\subsection{Monte Carlo Simulation}

\begin{figure*}[t]
    \centering
    \subfloat[GOSPA error for small Monte Carlo scenario with 8 objects and 5 sensors and $\sigma=1.0$.\label{fig:mc_minimal_gospa_low_noise}]{\includegraphics[width=0.45\linewidth]{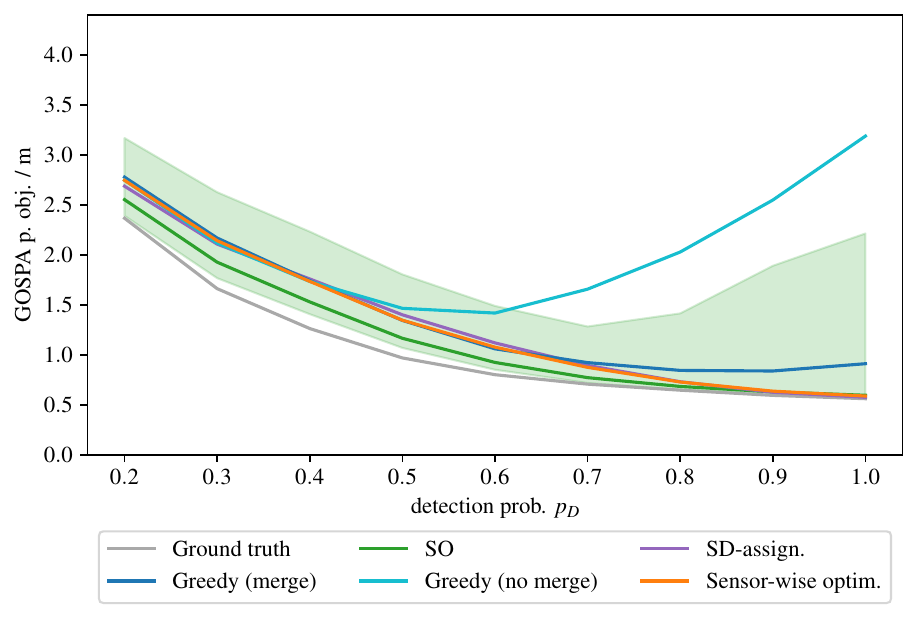}}
    \qquad
    \subfloat[GOSPA error for small Monte Carlo scenario with 8 objects and 5 sensors and $\sigma=2.0$.\label{fig:mc_minimal_gospa_high_noise}]{\includegraphics[width=0.45\linewidth]{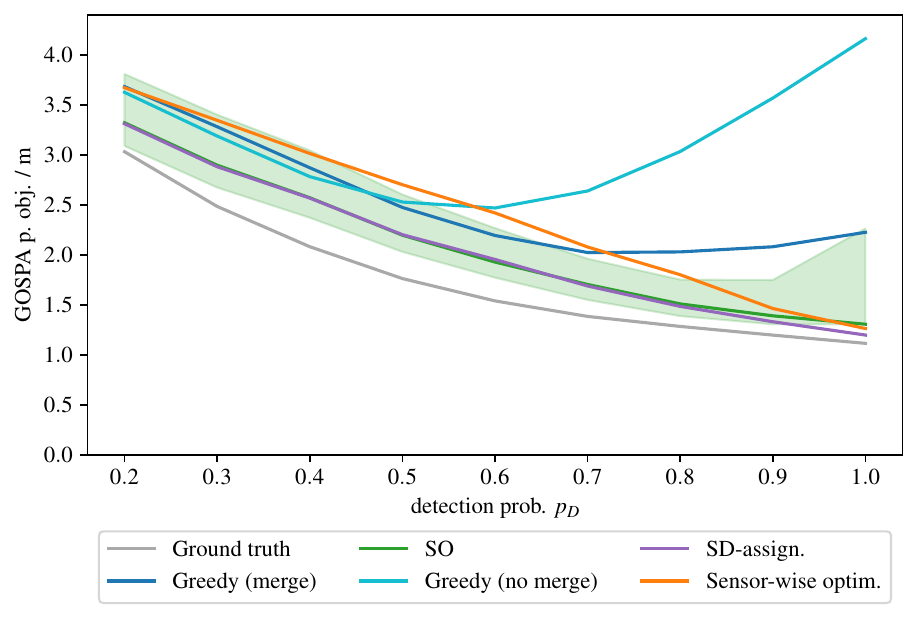}}
    \caption{\ac{gospa} error for small Monte Carlo scenario, two noise variants and different detection probabilities. For SO the GOSPA error of the 5 best hypotheses is shown as the shaded green area.}
    \label{fig:mc_minimal_gospa}
\end{figure*}

\begin{figure*}[t]
    \centering
    \subfloat[GOSPA error for big Monte Carlo scenario with 20 objects and 12 sensors and $\sigma=1.0$.\label{fig:mc_medium_gospa_low_noise}]{\includegraphics[width=0.45\linewidth]{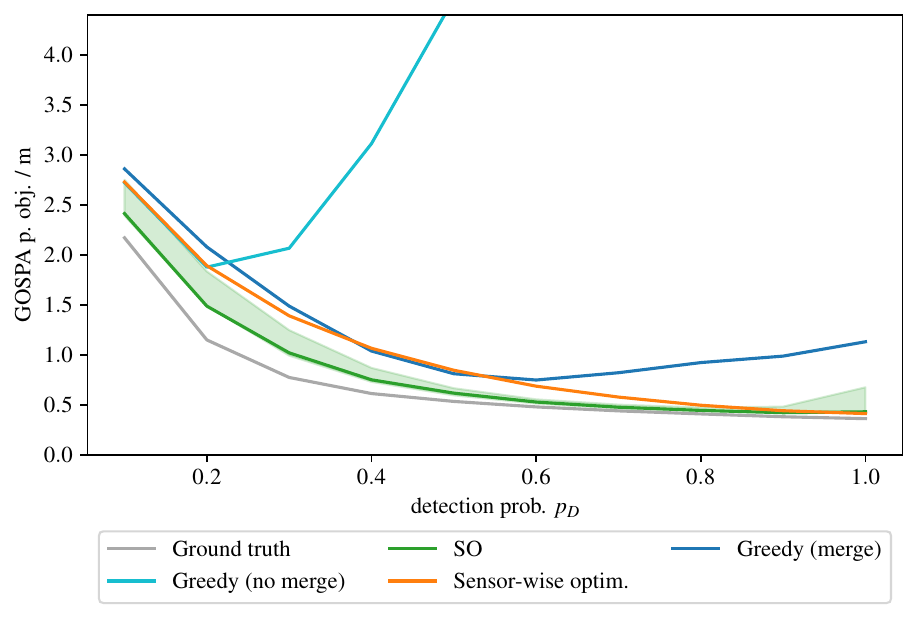}}
    \qquad
    \subfloat[GOSPA error for big Monte Carlo scenario with 20 objects and 12 sensors and $\sigma=2.0$.\label{fig:mc_medium_gospa_high_noise}]{\includegraphics[width=0.45\linewidth]{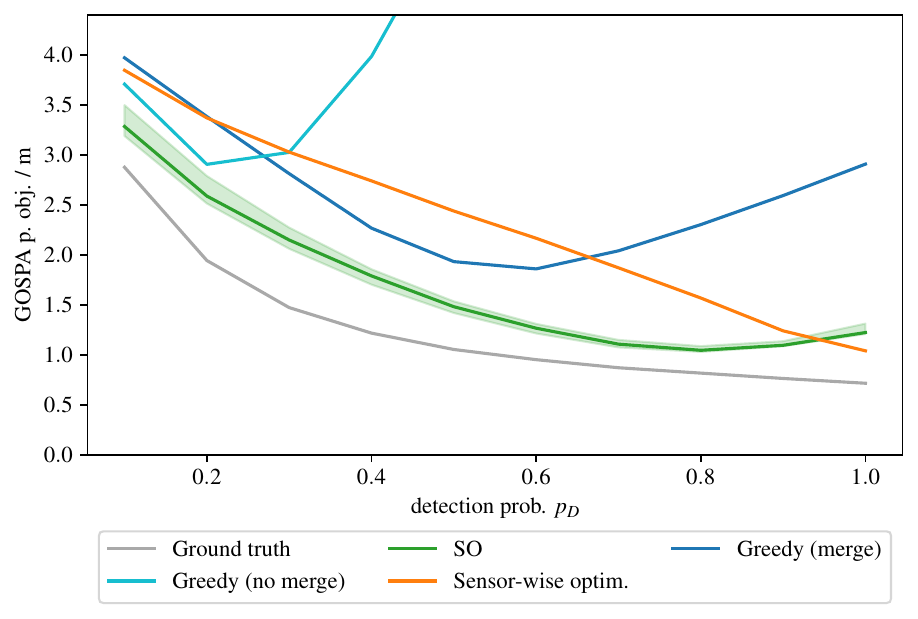}}
    \caption{\ac{gospa} error for big Monte Carlo scenario, two noise variants and different detection probabilities. For SO the GOSPA error of the 10 best hypotheses is shown as the shaded green area.}
    \label{fig:mc_medium_gospa}
\end{figure*}

The averaged \ac{gospa} error for different detection probabilities as well as two noise variants is shown in Fig.~\ref{fig:mc_minimal_gospa} for the small Monte Carlo scenario and in Fig.~\ref{fig:mc_medium_gospa} for the big Monte Carlo scenario.
The ground truth results are obtained by fusing the tracks that originated from the same object. As the tracks are noisy, the fused estimates using the correct association still yield a nonzero error,  which is higher with higher noise in both scenarios but decreases with higher detection probability.

In low noise \ac{so} performs best for all detection probabilities followed by sensor-wise optimization and SD-assign, which yield very similar results in the small scenario. Greedy (no merge) yields good results for low detection probabilities but the error increases drastically for $p_D>0.5$ in the small and $p_D>0.2$ in the big scenario. Greedy (merge) performs well in the small scenario, but the error increases slightly for $p_D>0.7$. In the big scenario it has the highest error for $p_D\leq0.2$, comparable errors to sensor-wise optimization and then increasing errors for $p_D>0.6$. 

In the high noise setting all errors increased. In the small scenario SD-assign has the best results. \ac{so} yields the same results but the error increases for very high detection probabilities ($p_D\geq 0.9$). In the big scenario \ac{so} yielded the best results except for very high detection probabilities. The performance of sensor-wise optimization decreased compared to the low noise scenarios but yields good results for very high detection probabilities in the small and big evaluation. Both Greedy variants behave similarly as in the low noise scenarios but with higher errors. 

The shaded green area highlighting the GOSPA error of the five or ten best SO associations shows that for very high $p_D$ these can have a much higher GOSPA error, but otherwise can even have a lower GOSPA error than the association with the highest likelihood.

\begin{figure}[t]
    \centering
    \includegraphics[width=0.95\linewidth]{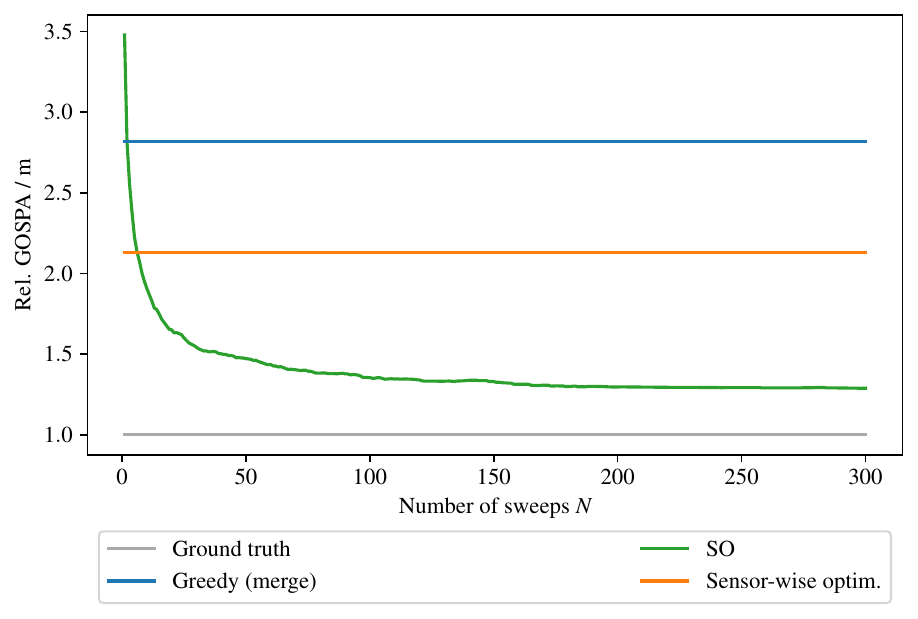}
    \caption{Convergence analysis of big Monte Carlo scenario with $p_D=0.8$ and $\sigma=2.0$ showing the relative \ac{gospa} error over different number of sweeps $N$.\label{fig:weight_convergence}}

\end{figure}

In Fig.~\ref{fig:weight_convergence} the relative \ac{gospa} error over the number of sweeps is shown. The relative \ac{gospa} error is computed as the \ac{gospa} error divided by the \ac{gospa} error of the correct association. Hence, the ground truth association has a relative \ac{gospa} error of $1$. As all but SO only compute one association, their error is constant over the number of sweeps. The error of \ac{so} decreases quickly and is lower than the other algorithms after only a couple of sweeps. For $50 < N < 200$ the relative \ac{gospa} error decreases only slightly and for $N\geq200$ the error remains constant.

\subsection{Collective Perception Scenario}

\begin{figure*}[t]
    \centering
    \subfloat[Full communication. \label{fig:artery_high_full}]{\includegraphics[width=0.45\linewidth]{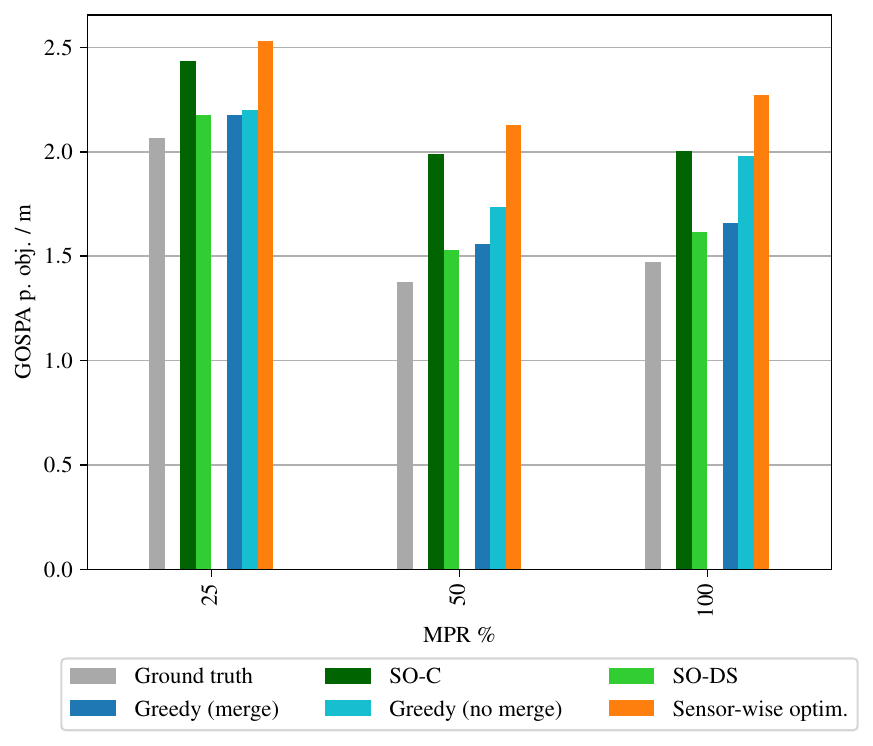}}
    \qquad
    \subfloat[ETSI rules. \label{fig:artery_high_etsi}]{\includegraphics[width=0.45\linewidth]{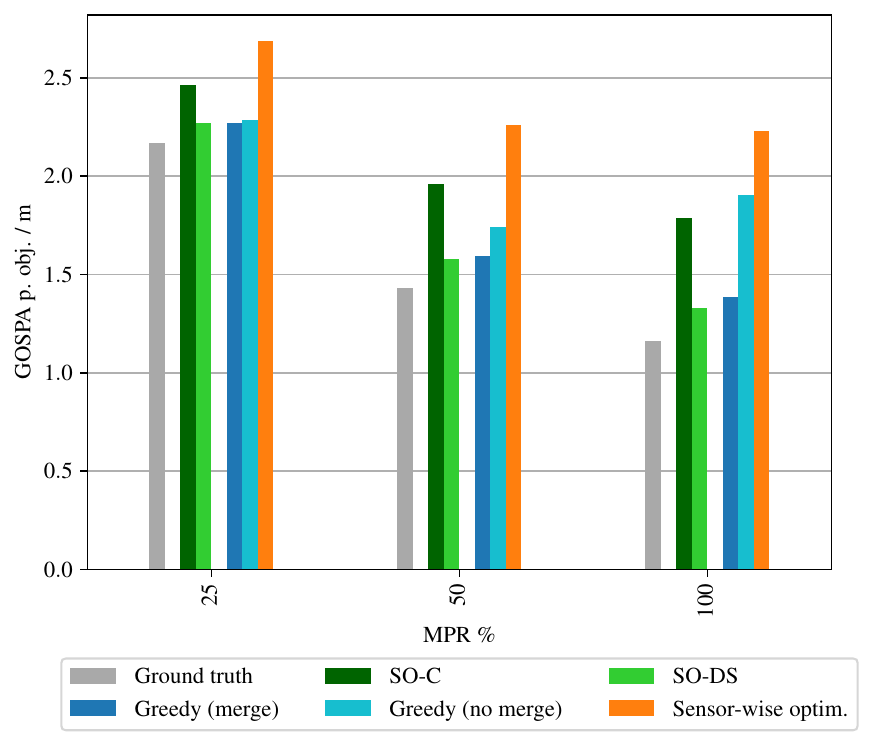}}
    \caption{Detailed \ac{gospa} results for the collective perception scenario. The results for full communication are shown in Fig.~\ref{fig:artery_high_full} and for the ETSI rules in Fig.~\ref{fig:artery_high_etsi}. Both are given for three market penetration rates.
    \label{fig:artery_high}}

\end{figure*}

Fig.~\ref{fig:artery_high} shows the \ac{gospa} results of the collective perception scenario. As the number of objects varies over time, we divided the \ac{gospa} error by the number of objects in each time step and then averaged over all time steps. Fig.~\ref{fig:artery_high} shows the errors for both communication scenarios (full communication and ETSI rules) and three different market penetration rates, i.e., the fraction of vehicles that is equipped with sensors. 

\ac{so-ds} yields the lowest error overall, followed by Greedy (merge), which always has a slightly higher error. 
Then follow Greedy (no merge), \ac{so-c} and sensorwise optimization with higher errors. 
Overall the error decreases with increasing MPR, including the error of the ground truth association. The only exception being the full communication scenario with {MPR~$=\SI{100}{\percent}$}, which yields a higher ground truth error than {MPR~$=\SI{50}{\percent}$}. The lowest error overall is achieved in the {MPR~$=\SI{100}{\percent}$} scenario with ETSI rules.

An evaluation of the PDR can be found in Tab.~\ref{tab:pdr}. In both scenarios there is no packet loss for the low MPR, but packet loss increases for the higher MPRs in both scenarios. However, the PDR for the ETSI rules is higher than for the full communication scneario and  still at 0.99 for MPR $=\SI{50}{\percent}$.

\subsection{Likelihood comparison}
\begin{figure}[t]
    \centering
    \includegraphics[width=0.95\linewidth]{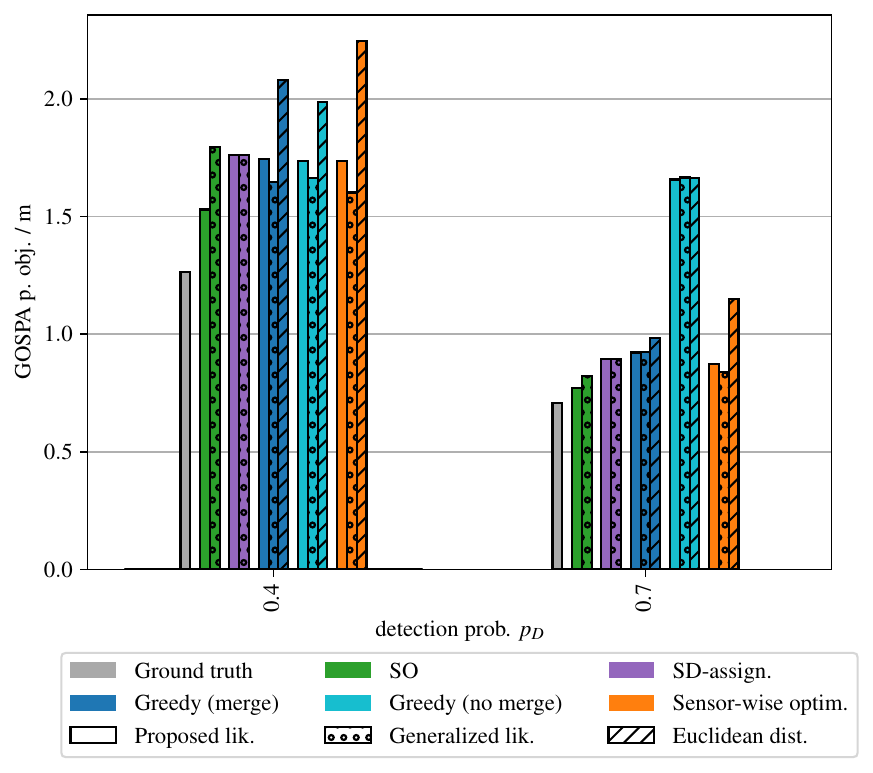}
    \caption{GOSPA results for different likelihoods in the small Monte Carlo scenario with $\sigma=1.0$.\label{fig:lik_comp_mc}}

\end{figure}

\begin{figure}[t]
    \centering
    \includegraphics[width=0.95\linewidth]{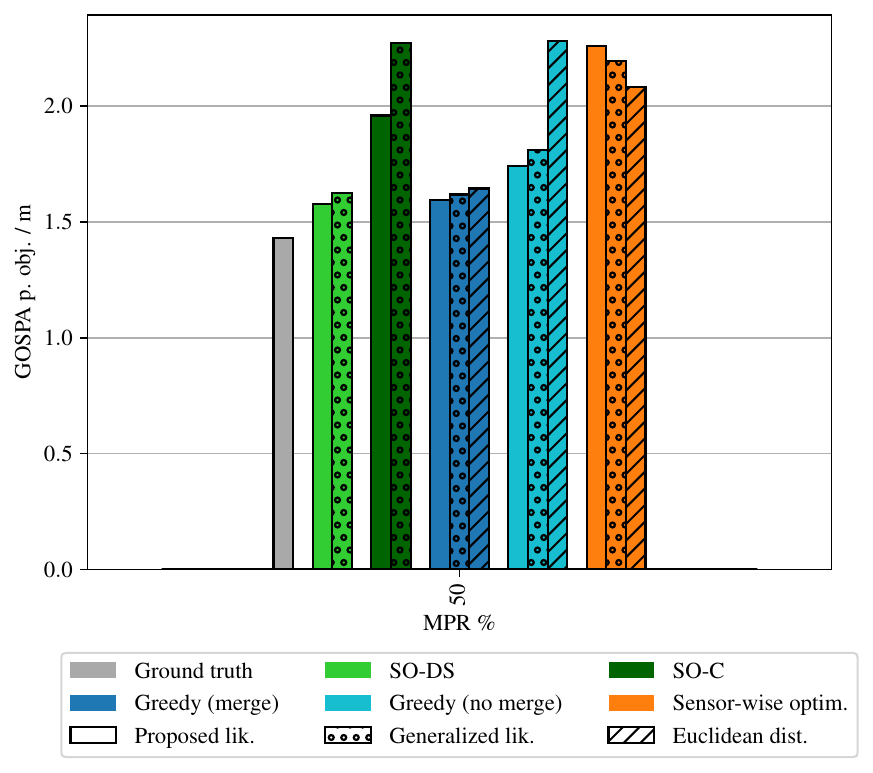}
    \caption{GOSPA results for one exemplary collective perception scenario with ETSI communication rules for different likelihoods.\label{fig:lik_comp_cp}}

\end{figure}

Fig.~\ref{fig:lik_comp_mc} contains the \ac{gospa} results of the small MC scenario with two detection probabilities with different objective functions. We applied our proposed likelihood as well as the generalized likelihood. For both Greedy variants and sensor-wise optimization we also applied the Euclidean distance. 

\ac{so} performs better with our proposed likelihood, especially for the lower detection probability. SD-assign yields the same results with both likelihoods. For the Greedy variants the Euclidean distance yielded the highest errors, while the generalized likelihood performed best overall with similar results for high detection probabilities for our proposed likelihood. 

Fig.~\ref{fig:lik_comp_cp} shows the averaged \ac{gospa} error of the collective perception scenario with ETSI rules and {MPR $=\SI{50}{\percent}$}. We again applied the same objective functions as in Fig.~\ref{fig:lik_comp_mc}. In the collective perception scenario overall our proposed likelihood yielded the lowest errors followed by the generalized likelihood and the Euclidean distance. The only exception is sensor-wise optimization which performed best with the Euclidean distance and worst with our proposed likelihood.

\section{DISCUSSION}
\label{sec:discussion}

In this section we discuss the results of our analyses of the Monte Carlo simulation and the collective perception scenario as well as the complexity of the algorithms and the choice of experiments.

\subsection{Monte Carlo simulation}
It can be seen that most algorithms were capable of computing good associations for different detection probabilities with low noise, but the association becomes more difficult with higher noise.
The merging step greatly improved the Greedy algorithm when three or more sensors are present. In the small Monte Carlo scenario with $p_D=0.6$ the expected number of tracks per object is 3 and the performance of Greedy (no merge) degraded for $p_D\geq0.6$ as clusters are often split into multiple small clusters and those clusters cannot be merged, which we also showed in~\cite{wolfTracktotrackAssociationBased2023}. 
On the other hand the merging step might be a disadvantage when only few sensors are present, see Fig.~\ref{fig:mc_minimal_gospa_high_noise} and~\ref{fig:mc_medium_gospa_high_noise}, where Greedy (no merge) performed better than Greedy (merge) for low detection probabilities. This could be caused by merging neighboring clusters when all tracks stem from different sensors. This effect can only partially be mitigated by reducing the gating threshold, as two objects can be closely spaced.

The sensor-wise optimization depends on the order the sensors are considered, which can lead to wrong associations for closely spaced objects. On the other hand it is not prone to generate too many clusters, as it aims to assign each track to the closest cluster if it is within the gating distance, which explains the good performance in very high detection probabilities. 

SD-assign yielded good results for the low noise scenario and very good results for the high noise scenario. However, the construction of the cost matrix requires to compute the likelihood of every possible cluster which is intractable for scenarios with more objects or sensors. 

\ac{so} yielded the overall best results. Unlike the Greedy variants and sensor-wise optimization it utilizes the multidimensional likelihood and not only pairwise likelihoods. In contrast to SD-assign it is applicable for bigger scenarios. It is the only approach that considers multiple hypotheses for the association as it is sampling-based. The convergence analysis showed that only a few sweeps are sufficient to achieve a good association. However, computing many samples does not necessarily yield better associations. 
Especially for very high $p_D$ SO did not always find the best association. SO can get stuck in local optima, especially with very high $p_D$ as the likelihood of a smaller cluster is much smaller and merging two clusters is often not possible due to the sensor constraint. Then, if a good association is reached it is very unlikely to sample the steps necessary to get to a better association.
The shaded green areas indicated that using only one association, even if it is the one with the highest possible likelihood might not yield the best fused result regarding the GOSPA error. 

\subsection{Collective perception scenario}
Unlike in the Monte Carlo simulation the correct detection probability in the collection perception scenario is not known and the tracks are generated by trackers. As \ac{so-c} did not perform well in most of the scenarios, it can be said that a uniform approximation of the detection probability, which further depends on the estimation of the observable area, is not suitable. \ac{so-ds}, which incorporates the sensor range is more appropriate as it yielded the best results in all scenarios. Currently the detection probability is assumed to be constant within the sensor range. With more data a more accurate approximation could be achieved. 
It has to be noted that both Greedy variants and the sensor-wise optimization only use the spatial likelihood as they are based on pairwise likelihoods and thus the detection probability does not have to be estimated. 

The overall error decreased with higher MPR, as more sensors are present (see also Tab.~\ref{tab:artery_stats}) and fewer objects are misdetected. Using the ETSI rules on one hand decreased packet loss, although it could not be completely mitigated for MPR~$\SI{100}{\percent}$. On the other hand it greatly increased the number of sensors as the acp{cpm} were not lost and furthermore saved over several time steps, which complicates the association. 

\subsection{Likelihood}
Our proposed likelihood greatly improved \ac{so}, especially in scenarios with small clusters, i.e., when the detection probability is low or there are not many sensors. Then, the uncertainty of the cluster center has a greater impact. However, for SD-assign, which aims to optimize the joint association instead of evaluating actions for single tracks or clusters, both likelihoods yielded the same results. 

Both Greedy variants benefited from using a likelihood instead of the Euclidean distance. In some cases our likelihood yielded better results and in others the generalized likelihood. However, the Greedy variants and sensorwise optimization heavily depend on the distance threshold $d_t$, which has to be chosen carefully. If it is chosen too high, tracks from different objects can be associated or in the case of Greedy (merge) clusters from different objects can be merged. If it is chosen too low, objects are split into different clusters. The distance thresholds in this paper have been evaluated experimentally but no fine-grained optimization has been conducted and thus the choice of distance threshold can have a bigger impact than the choice of likelihood. 

\subsection{Complexity}

The Greedy approach, with or without merging, has the lowest complexity as it is quadratic in the number of tracks. An upper bound for the number of tracks is $\numsensors\cdot\numobjects$. In that case both Greedy variants and \ac{so} are quadratic each in the number of sensors and objects. \ac{so} is furthermore linear in the number of sweeps. On the contrary, SD-assign and sensor-wise optimization are linear in the number of sensors and cubic in the number of objects, which make them less suitable for scenarios with many objects. SD-assign furthermore requires to compute a high-dimensional cost matrix, which is intractable already for moderate problem sizes. 

\ac{so} has the lowest complexity to compute just one association, however, just one association is not sufficient. Fig~\ref{fig:weight_convergence} shows that \ac{so} converges quickly and not many sweeps are necessary to achieve a good association result. Furthermore, the computation of $\mat{p_A}$ could be parallelized and computation time can be improved using caching.

\subsection{Scenarios}
The Monte Carlo simulation has the advantage that all parameters are known exactly. However, it does not represent realistic tracking scenarios as all tracks have the same covariance and objects are randomly placed. The collective perception scenario is more realistic regarding object placement, the tracks stem from a tracking algorithm and the detections are based on sensor range and occlusions. Yet, even a realistic simulation has its limitations, as the same tracker is applied in all intelligent vehicles and we only use point measurements.

\section{CONCULUSION AND FUTURE WORK}
\label{sec:conclusion}
In this paper we presented a novel \ac{t2ta} algorithm based on stochastic optimization. It yielded the overall best results in the Monte Carlo simulation with varying detection probabilities and noise parameters as well as in the collective perception scenario with different communication strategies. It is suitable for settings with many sensors as it only has a quadratic complexity in the number of tracks and does not depend on parameter tuning like the Greedy variants. Furthermore, it is the only approach generating multiple associations that better represent ambiguous data association situations with closely spaced objects and that could be used in a multi-hypothesis track fusion approach. 

The proposed likelihood improved the performance of \ac{so}, especially in scenarios with low detection probability as the uncertainty of the estimated cluster center has a greater impact for small clusters. Using a likelihood instead of the Euclidean distance also has a great impact on the Greedy variants, although the generalized likelihood yielded better results than our proposed likelihood in some cases. However, the Greedy variants and also sensorwise optimization depend on distance threshold that has to be adjusted to the applied objective function and evaluation scenario. 

The collective perception scenario showed that it is not realistic to assume full communication as this congests the communication channel and leads to packet loss. The ETSI rules that are designed to reduce channel load increase the complexity of the \ac{t2ta} problem as the number of sensors per time step is higher and the tracks that have to be associated can be older and have to be propagated to the current time. Furthermore, the assumption of a constant detection probability does not hold as the sensors' field of view is not identical. 

In future work we want to investigate the declining performance of \ac{so} in scenarios with many sensors and very high detection probability and if this could be mitigated by introducing other actions in the \ac{so} approach. To address the computational complexity in very large scenarios one direction of future work could be to combine the \ac{so} approach with a clustering approach as a two-step process, where \ac{so} is only used, when objects are not easily separable, as it was done similarly for  SD-assign in~\cite{chummunFastDataAssociation2001}.

\bibliographystyle{IEEEtran}
\bibliography{T2TA}

\end{document}

%% file: SDF_Header.tex
\usepackage{amsfonts}
\usepackage{amssymb}
\usepackage{mathrsfs}
\usepackage{amsmath}
\usepackage{mathtools}
\usepackage{url}

\usepackage{tikz}
\usepackage{pgfplots}
\DeclareUnicodeCharacter{2212}{−}
\usepgfplotslibrary{groupplots,dateplot}
\usetikzlibrary{patterns,shapes.arrows}
\pgfplotsset{compat=newest}

\usepackage{xargs}
\usepackage{color}

\definecolor{tcolor}{RGB}{0,60,104}
\usepackage{pgfplots}

\newcommand{\rvec}[1]{\ensuremath{{\boldsymbol{{#1}}}}}
\newcommand{\mat}[1]{{\ensuremath{{\mathbf{#1}}}}}

\DeclareMathOperator*{\argmin}{arg\,min}

\DeclareMathOperator{\diag}{diag}

\newcommand{\Fig}[2][KeinUnterVerweis]
{\ifthenelse{\equal{#1}{KeinUnterVerweis}}%
{Fig.~\ref{#2}}
{Fig.~\hbox{\ref{#2} #1}}%
}%

\newcommand{\jointasso}{\rvec{\theta}}
\newcommand{\jointassoit}[1]{{\jointasso^{(#1)}}}
\newcommand{\clusterlist}{\mathcal{C}}
\newcommand{\cluster}[1]{{\mat{C}_{#1}}}

\newcommand{\trackset}{\mathcal{T}}
\newcommand{\numtracks}{{N_{\trackset}}}
\newcommand{\sensorset}{\mathcal{S}}
\newcommand{\numsensors}{{N_{\sensorset}}}
\newcommand{\numobjects}{{N_O}}
\newcommand{\state}[1]{\mat{x}_{#1}}
\newcommand{\cov}[1]{\mat{P}_{#1}}
\newcommand{\sensorpos}[1]{\mat{s}_{#1}}
\newcommand{\clustercenter}{\overline{\mat{x}}^c}
\newcommand{\clustercov}{\overline{\mat{P}}^c}

\DeclareMathOperator{\randchoice}{randChoice}
